\documentclass[pra,nofootinbib,twocolumn]{revtex4-1}

\usepackage{amsmath,amsfonts,bm,dsfont,mathrsfs,url}
\usepackage{enumitem}

\newcommand{\mc}{\mathcal}

\newcommand{\ket}[1]{|#1\rangle}
\newcommand{\bra}[1]{\langle#1|}

\newcommand{\nrm}[1]{\| #1 \|}

\newtheorem{corollary}{Corollary}
\newtheorem{theorem}{Theorem}

\begin{document}


\title{Relativistically Invariant Markovian Dynamical Collapse Theories Must Employ Nonstandard Degrees of Freedom}


\author{Wayne C. Myrvold}
\affiliation{Department of Philosophy, The University of Western Ontario}
\email{wmyrvold@uwo.ca}


\date{\today}

\begin{abstract}
 The impossibility of an indeterministic evolution  for standard relativistic quantum field theories, that is, theories in which all fields satisfy the condition that the generators of spacetime translation have spectrum in the forward light-cone, is demonstrated.  The demonstration proceeds by arguing that a  relativistically invariant theory must have a stable vacuum, and then showing that  stability of the vacuum, together with the requirements imposed by relativistic causality, entails deterministic evolution, if all degrees of freedom are standard degrees of freedom.
\end{abstract}


\maketitle

\section{Introduction} Dynamical collapse theories replace the deterministic, unitary evolution of the quantum state with a stochastic evolution, with the aim of providing a dynamics that  suppresses superpositions of macroscopically distinguishable states. Well-known examples are the GRW model \cite{GRW} and the Continuous Spontaneous Localization model (CSL) \cite{Pearle1989,GPR90} which, unlike the GRW model, respects symmetrization and antisymmetrization requirements on wavefunctions for identical particles. For reviews, see refs. \cite{BGSurvey,CollapseReview2013}.

There is a long history of attempts to extend the CSL model to relativistic quantum field theories.\footnote{There is also an extension of the GRW theory to a relativistic context \cite{DoveDiss,DoveSquires96,TumulkaRGRW}.  This model is restricted to a theory involving a fixed, finite number of noninteracting Dirac particles.  As any relativistic quantum theory with interactions must be a quantum field theory, this article is concerned, not with theories of this sort, but with the prospects of a  collapse theory for a relativistic quantum field theory.}  Straightforward extensions of the model to the context of relativistic QFT have the physically unacceptable feature of an infinite rate of energy production per unit volume from the vacuum \cite{Pearle62Years,GGP1990,GGP1991,PearleTachyonic,AdlerBrun2001,NicroRimini,BGSurvey}.  Relativistic collapse models formulated by Bedingham  \cite{BedinghamRSRM,BedinghamRSRD} and Pearle \cite{Pearle2015} achieve a stable vacuum, but they do so by invoking a nonstandard quantum field with no intrinsic dynamics. The purpose of this article is to show that a move along these lines is necessary; the constraints that relativity places on a stochastic theory make it impossible to construct a  Markovian stochastic collapse theory in Minkowski spacetime employing only standard degrees of freedom.

We will first review the general framework of quantum stochastic dynamical semi-groups, and then consider the constraints imposed by relativity.

\section{Stochastic quantum state evolution} A \emph{quantum dynamical semi-group} is a one-parameter family of trace-preserving linear mappings $\{\Sigma_t, t \geq 0 \}$ on the set of density operators, continuous in trace norm, such that
\begin{itemize}
\item $\Sigma_0(\rho) = \rho$;
\item For all $t, s$, $\Sigma_s (\Sigma_t (\rho))  = \Sigma_{t+s}(\rho)$.
\end{itemize}
There are good reasons to hold that any physically reasonable evolution of the quantum state must be given by a completely positive mapping on the space of quantum states.  Accordingly, we will  take the elements of the quantum dynamical semi-group  to be completely positive.  They therefore can be represented in Kraus form, via operators $\{K_i(t)\}$, such that
\begin{equation}
\rho(t + s) =  \sum_i K_i(s) \, \rho(t) \, K^\dag_i(s),
\end{equation}
with
\begin{equation}\label{tracep}
\sum_i K_i^\dag(s) K_i(s) = \mathds{1}.
\end{equation}
We can use this framework to define a general schema for Markovian stochastic state evolution.   Given a state vector $\ket{\psi(t)}$, for any $s > 0$ such an evolution should specify a set of alternatives for $\ket{\psi(t + s)}$, and a probability measure over the set of alternatives.  We consider first the case of a discrete set of alternatives. We assume, for every $s > 0$, a set of operators  $\{K_i(s) \}$, satisfying (\ref{tracep}), such that, for some $i$,
\begin{equation}
\ket{\psi(t + s)} = K_i(s) \ket{\psi(t)}/\nrm{K_i(s) \ket{\psi(t)}},
\end{equation}
with probabilities
\begin{equation}
p_i = \nrm{K_i(s) \ket{\psi(t)}}^2.
\end{equation}
The state, at any time, is, therefore, a pure state, but \emph{which} pure state obtains at a later time is not uniquely determined by the state at an earlier time.  In addition to the pure-state density operator representing the state at $t + s$,  there will, for $s > 0$, be a mixed state density operator $\bar{\rho}(t + s)$ , which may be called the \emph{ensemble}, or \emph{mean} density operator,  which is a weighted average of the various possibilities for $\rho(t + s)$, given the state at $t$, weighted by their respective probabilities.
\begin{equation}
\bar{\rho}(t + s) = \sum_i K_i(s) \, \rho(t) \, K^\dag_i(s).
\end{equation}

We can also consider stochastic processes of a more general sort.  Let $\langle \Gamma, \mc{F}, \mu \rangle$ be a measure space.  Suppose that  we have a family of operators $\{K_\gamma\}$, for $\gamma \in \Gamma$,  such that
\begin{equation}
\int_\Gamma d\mu(\gamma) \, K^\dag_\gamma K_\gamma  = \mathds{1}.
\end{equation}
These can serve as the operators that induce our state transitions; we will call them, accordingly \emph{evolution operators}.

This gives us a rather general schema for a stochastic process in a Hilbert space.    For the moment, we will presume that we have a unique global time function; extension to a relativistic spacetime will be considered in the next subsection.

Any theory satisfying the following conditions will give us stochastic evolution of the state vector.
\begin{enumerate}
\item   For each time interval $[t, t + s]$, there is a measure space  $\langle \Gamma_s, \mc{F}_s, \mu_s \rangle$.  The elements of $\Gamma_s$  are to be thought of as indexing possible events in that interval.  These measure spaces must satisfy the conditions that, for $s' > s$,
    \begin{enumerate}
    \item $\Gamma_s \subseteq \Gamma_{s'}$,
     \item $\mc{F}_s \subseteq \mc{F}_{s'}$, and
     \item  $\mu_s$ is the restriction of $\mu_{s'}$ to $\mc{F}_s$.
    \end{enumerate}
\item For each $s$, there  is a measurable function that maps $\gamma \in  \Gamma_s$ to an operator $K_\gamma(s)$, such that
\[
\int_{\Gamma_s}d \mu_s(\gamma) \, K_\gamma^\dag(s) \, K_\gamma(s)    = \mathds{1}.
\]
\item The state vector $\ket{\psi(t + s)}$ is a random variable, defined as follows.  For some $\gamma \in \Gamma_s$,
\[
\ket{\psi(t + s)} = K_\gamma(s) \ket{\psi(t)}/\nrm{K_\gamma(s) \ket{\psi(t)}},
\]
with the probability rule,
\[
Pr(\gamma \in F) = {\int_F d \mu_s(\gamma) \, \|  K_\gamma(s) \ket{\psi(t)} \|^2}.
\]
\end{enumerate}

On a theory of this sort, given an initial condition $\ket{\psi(t_0)}$, for any later time $t_1 = t_0 +  s$ there will be an actual state vector $\ket{\psi(t_1)}$, and its corresponding pure-state density operator
\begin{equation}
\rho(t_1) = \ket{\psi(t_1)} \bra{\psi(t_1)}.
\end{equation}
The \emph{ensemble} density operator is
\begin{equation}\label{KrausEnsemble}
\bar{\rho}(t_1) =  \int_{\Gamma_{s}}d\mu_s(\gamma) \, K_\gamma(s) \rho(t_0) K^\dag_\gamma(s).
\end{equation}

\section{Stochastic state evolution in Minkowski spacetime}
\subsection{Framework} It is useful to work within what may be called the \emph{stochastic Tomonaga-Schwinger picture}.  The usual Tomonaga-Schwinger picture is an adaptation of the interaction picture to a relativistic spacetime. In this picture, one writes the Lagrangian density as a sum of a free-field Lagrangian and a term incorporating interactions:
\begin{equation}
\mc{L}(x) = \mc{L}_0(x) + \mc{L}_I(x).
\end{equation}
The operators employed are solutions of the free-field equations, and one associates, with each spacelike Cauchy surface $\sigma$ (whether flat or not), a statevector $\ket{\psi(\sigma)}$.  Evolution from a surface $\sigma$ to another, $\sigma'$,  differing by an infinitesimal  deformation about a point $x$, satisfies the \emph{Tomonaga-Schwinger equation} \cite{Tomonaga46,Schwinger48}:
\begin{equation}
i \hbar  c\, \frac{\delta \ket{\psi(\sigma)}}{\delta \sigma(x)} = \mc{H}_I(x) \ket{\psi(\sigma)}.
\end{equation}
Integration of this equation yields, for any Cauchy surfaces $\sigma$, $\sigma'$, a  unitary mapping from $\ket{\psi(\sigma)}$ to $\ket{\psi(\sigma')}$.

In the \emph{stochastic Tomonaga-Schwinger picture}  \cite{Pearle62Years,GGP1990,AdlerBrun2001,BGSurvey}, we employ Heisenberg-picture operators that are solutions of the standard equations, for free or interacting fields.  The additional terms that implement evolution from one Cauchy surface to another are those that are responsible for collapse.  The schema for stochastic evolution outlined in the previous section is readily adapted to this setting.
\begin{enumerate}
\item   For each pair of Cauchy surfaces $\sigma$, $\sigma'$, with $\sigma'$ nowhere to the past of $\sigma$, let $\delta$ be the spacetime region between them.  For each such $\delta$,  there is a probability space  $\langle \Gamma_\delta, \mc{F}_\delta, \mu_\delta \rangle$, where the elements of $\Gamma_\delta$ are to be thought of a indexing the possible evolutions  that can occur between $\sigma$ and $\sigma'$.  These measure spaces must satisfy the conditions that, for $\delta \subseteq \delta'$,
    \begin{enumerate}
    \item $\Gamma_\delta \subseteq \Gamma_{\delta'}$,
     \item $\mc{F}_\delta \subseteq \mc{F}_{\delta'}$, and
     \item  $\mu_\delta$ is the restriction of $\mu_{\delta'}$ to $\mc{F}_\delta$.
    \end{enumerate}
\item For each $\delta$, there  is a measurable function  that maps $\gamma \in  \Gamma_\delta$ to an operator $K_\gamma(\delta)$, such that
\[
\int_{\Gamma_\delta}  d \mu_\delta(\gamma) \,  K_\gamma^\dag(\delta) \, K_\gamma(\delta)   = \mathds{1}.
\]
\item The state vector $\ket{\psi(\sigma')}$ is a random variable, such that, for some $\gamma \in \Gamma_\delta$,
\[
\ket{\psi(\sigma'} = K_\gamma(\delta) \ket{\psi(\sigma)}/\nrm{K_\gamma(\delta) \ket{\psi(\sigma)}},
\]
with the probability rule,
\[
Pr(\gamma \in F) = {\int_F d\mu_\delta(\gamma) \, \|  K_\gamma(\delta) \ket{\psi(\sigma)} \|^2}.
\]
\end{enumerate}
Consider two Cauchy surfaces, $\sigma$, $\sigma'$, which coincide everywhere except on the boundaries of two bounded regions $\delta_1$, $\delta_2$.  We must have a unique stochastic evolution from $\sigma$ to $\sigma'$, through $\delta_1 \cup \delta_2$, and this should coincide with the composition of the evolution through $\delta_1$ and the evolution through $\delta_2$, in either order.  The necessary and sufficient condition for this is,
\begin{enumerate}[resume]
\item \label{LocEv} Evolution operators  corresponding to spacelike separated regions commute.
\end{enumerate}
Moreover, for computing probabilities pertaining to experiments performed on the overlap of  $\sigma$ and $\sigma'$, it should not matter whether $\rho(\sigma)$ or $\bar{\rho}(\sigma')$ is employed.  The necessary and sufficient condition for this is \cite{FixedPoints},
\begin{enumerate}[resume]
\item \label{NoSig} Evolution operators pertaining to a region $\delta$ commute with all operators representing observables at spacelike separation from $\delta$.
\end{enumerate}
The conditions \ref{LocEv}, \ref{NoSig} ensure compatibility with relativistic causal structure.

\subsection{Necessity for a stable vacuum}  A recurring difficulty that has arisen in attempts to create a version of CSL adapted to a relativistic quantum field theory is divergence of energy produced from the vacuum \cite{Pearle62Years,AdlerBrun2001,BGSurvey}.  The problem is that, if there is any particle production from the vacuum at all, it cannot be kept finite \cite{PearleTachyonic,Pearle2015}.  The vacuum state is invariant under the Poincar\'e group.  Therefore, if the theory produces  excitations from the vacuum, the probability distribution over such excitations must be invariant under spacetime symmetries, as there is nothing in the state that could be used to break the symmetry. This requires a probability measure over possible excitations that is invariant under the Poincar\'e group. The difficulty is the absence of finite invariant measures over the space of possible excitations.

This can be illustrated by the simplest case, that of a free theory.  If the theory assigns  a nonzero probability to producing an excitation that has momentum $k$ in a certain volume $\Delta$ of the mass-shell, it must assign the same probability to any Lorentz boost of $\Delta$.  The unique (up to a multiplicative constant) nonzero invariant measure on the mass-shell for mass $m$ assigns, to any set $\Delta$, measure
\begin{equation}
\omega(\Delta) = \int_\Delta \frac{d^3 \mathbf{k}}{\sqrt{\mathbf{k}^2 + m^2}}
\end{equation}
This diverges when extended to the entire mass-shell.

This appears, in the theories proposed,  formally as infinite particle production per unit of time in any given volume of space. It  is a symptom of a deep  problem, one that cannot be renormalized away: the only probability distribution over excitations from the vacuum that does not break the Poincar\'e symmetry of the vacuum is one on which there is strictly zero probability for any excitation.  For this reason, a relativistic collapse theory must have a stable vacuum.   This, as we shall see in the next section, poses a difficulty.

\section{An impossibility result}  We will now show that the conditions on stochastic evolution outlined in the previous section cannot be achieved by a theory employing only \emph{standard} degrees of freedom, in a sense that we will now define.

We assume a quantum field theory with various ``field operators'' $\phi_\alpha(x)$ (actually operator-valued distributions, which yield operators when smeared with appropriate test functions),   obeying the usual bosonic or fermionic commutation or anticommutation relations at spacelike separation. We assume a unitary representation of the group of spacetime translations, with infinitesimal generators $P^\mu$, such that
\begin{equation}
\phi_\alpha(x + a) = e^{i P a /\hbar}  \, \phi_\alpha(x) \, e^{-i P a /\hbar}.
\end{equation}
The generators of spacetime translations are required to satisfy the \emph{spectrum condition}:
\begin{quote}
When $a$ is a future-directed timelike vector, the spectrum of $P a$ is in $\mathds{R}^+$.
\end{quote}
This ensures positivity of the energy, with respect to any reference frame.  We will call fields whose spacetime translations are generated by operators satisfying the spectrum condition \emph{standard fields}, and operators formed from them, \emph{standard operators}.  These include all fields and operators appearing in standard quantum field theories; the reason for the terminology is to distinguish them from nonstandard fields that have been introduced in the context of dynamical collapse theories \cite{PearleWays,Pearle2005,Pearle2008,BedinghamRSRM,BedinghamRSRD,Pearle2015}.

We assume a unique vacuum state that is the ground state of all fields that appear in the theory, standard or nonstandard.  If the vacuum is stable, it must be an eigenstate of the global interaction Hamiltonian,
\begin{equation}
 H = \int  d^3 \mathbf{x}  \: \mc{H}_I(t, \mathbf{x}).
\end{equation}
Lorentz invariance of the vacuuum requires that it be an eigenstate of all other generators of spacetime translations.  Therefore, for a stable, Lorentz-invariant vacuum, the state on each spacelike hyperplane will  represented by the same Tomonaga-Schwinger picture state vector $\ket{\Omega}$.  The state need not be an eigenstate of the local Hamiltonian densities $\mc{H}_I(x)$, and therefore, for nonflat hypersurfaces $\sigma$ (which are not related to a flat surface by a spacetime symmetry), the unchanging vacuum may be represented by a Tomonaga-Schwinger picture state vector $\ket{\psi(\sigma)}$ that is different from the vector $\ket{\Omega}$.

For any open spacetime region $R$, let $\mc{P}(R)$  be the \emph{polynomial algebra} of $R$, that is, the algebra of operators generated by operators of the form
\begin{equation}
A = \int d^4 x_1 \ldots d^4 x_n \: f(x_1, \ldots, x_n) \: \Pi_{i = 1}^n \phi_{\alpha_i}(x_i),
\end{equation}
with $f$ nonzero only when all of its arguments are in $R$.  These polynomial algebras include, of course, $\mc{P}(M)$, where $M$ is the whole of Minkowski spacetime. Let $\mathscr{H}_S(R)$ be the Hilbert space that is the  completion in norm of the set of vectors that result from operation on $\ket{\Omega}$ with operators in $\mc{P}(R)$.  The \emph{standard Hilbert space} of the theory,  $\mathscr{H}_S$, is $\mathscr{H}_S(M)$.

The Reeh-Schlieder theorem \cite{RS61}, which says that, for any $R$, any vector in  $\mathscr{H}_S$ can be approximated to arbitrary accuracy by applying elements of $\mc{P}(R)$ to the vacuum state $\ket{\Omega}$,  applies.
\begin{theorem} (Reeh-Schlieder).  For any open region $R$, $\mathscr{H}_S(R) = \mathscr{H}_S$.
 \end{theorem}
 An immediate corollary of the Reeh-Schlieder theorem is:
 \begin{corollary}
 For any operator $A$ and any open region $R$, if  $A$ either commutes or anticommutes with all standard fields $\phi_\alpha(x)$ with $x \in R$, then, if $A \ket{\Omega} = 0$, $A \ket{\Phi} = 0$ for all $\ket{\Phi} \in \mathscr{H}_S$.
 \end{corollary}
 This poses a difficulty for relativistic collapse theories, as there is a tension between the condition of stability of the vacuum and the requirements imposed by relativistic causality.  Let $\sigma_0$ be a spacelike hyperplane, and let  $\sigma_1$ be a spacelike hypersurface that coincides with $\sigma_0$  everywhere except for  a bounded region in which $\sigma_1$ lies to the future of $\sigma_0$. Let $\delta$ be the spacetime region between $\sigma_0$ and $\sigma_1$. Let $\{K_\gamma(\delta)\}$ be the set of operators that implement the  evolution through $\delta$ from $\ket{\psi(\sigma_0)}$ to  $\ket{\psi(\sigma_1}$.  Suppose that $\ket{\psi(\sigma_0)}$ is the  vacuum state. For some $\gamma$,
 \begin{equation}
 \ket{\psi(\sigma_1)} =   K_\gamma(\delta) \ket{\psi(\sigma_0)}/\nrm{K_\gamma(\delta) \ket{\psi(\sigma_0)}}.
 \end{equation}
 As mentioned, stability of the vacuum does not require that  $\ket{\psi(\sigma_1)}$ be the same vector as $\ket{\psi(\sigma_0)}$, as these are Tomonaga-Schwinger picture representations of the states along different hypersurfaces, and our collapse theory might include (as does the theory of \cite{BedinghamRSRM,BedinghamRSRD}) an interaction term $\mc{H}_I(x)$ such that  $\ket{\Omega}$ is not an eigenstate of $\mc{H}_I(x)$ but \emph{is} an eigenstate of the total Hamiltonian.   However, stability of the vacuum \emph{does} require that $\ket{\psi(\sigma_1)}$ and $\ket{\psi(\sigma_0)}$ be two Tomonaga-Schwinger representations of an unchanging unique vacuum state, and this means that, if  $\ket{\psi(\sigma_0)} = \ket{\Omega}$, then
 \begin{equation}
 \bar{\rho}(\sigma_1) = \int d\mu(\gamma) \: K_\gamma(\delta) \: \ket{\Omega} \bra{\Omega} \: K^\dag_\gamma(\delta)
 \end{equation}
is pure.

In order for the state  $\bar{\rho}(\sigma_1)$ to be pure, it must be the case that, for any   $\gamma$, $\gamma'$, (except, perhaps, for an exceptional set with total probability zero of being realized), $K_\gamma(\delta)$ and $K_{\gamma'}(\delta)$  map $\Omega$ onto the same ray in Hilbert space.  That is, for almost all $\gamma$, $\gamma'$, there must exist $c$ such that
\begin{equation}
K_{\gamma'}(\delta) \ket{\Omega} = c \, K_\gamma(\delta) \ket{\Omega},
\end{equation}
or,
\begin{equation}
(K_{\gamma'}(\delta) -  c \, K_\gamma(\delta)) \ket{\Omega} = 0.
\end{equation}
If, now, as required by relativistic causality, $K_{\gamma}(\delta)$ and $K_{\gamma'}(\delta)$ commute with all operators representing observables in any open region $R$ spacelike separated from $\delta$, it follows, from the above corollary to the Reeh-Schlieder theorem, that
\begin{equation}
(K_{\gamma'}(\delta) - c \, K_\gamma(\delta)) \ket{\Phi} = 0
\end{equation}
for all $\ket{\Phi} \in \mathscr{H}_S$.  That is, for almost all $\gamma, \gamma'$, for any state $\ket{\Phi}$ in the standard Hilbert space, $K_\gamma(\delta)$ and $K_{\gamma'}(\delta)$ map $\ket{\Phi}$ to vectors in the same ray.  This means that the evolution from $\ket{\psi(\sigma_0)}$ to $\ket{\psi(\sigma_1)}$ is deterministic (that is, $\bar{\rho}(\sigma_1)$ is pure), for \emph{any} initial state $\ket{\psi(\sigma_0)}$ in the standard Hilbert space $\mathscr{H}_S$.

What one wants, from a collapse theory, is, of course, indeterministic evolution; linear deterministic evolution of the quantum state vector will result in the sorts of superpositions of macroscopically distinguishable states that  dynamical theory are designed to suppress.  We have shown that this is impossible, within our framework of stochastic quantum state evolution, as long as the theory remains within the standard Hilbert space $\mathscr{H}_S$.

 As mentioned in the introduction, there are relativistic collapse models with  a stable vacuum.  These employ nonstandard degrees of freedom, so that the full Hilbert space of the theory goes beyond the standard Hilbert space $\mathscr{H}_S$.  The models of refs. \cite{BedinghamRSRM,BedinghamRSRD,Pearle2015} utilize a nonstandard field  introduced in ref. \cite{PearleWays}, called  an ``index'' or ``pointer'' field. This field  commutes with itself at any distinct points, whether spacelike or timelike separated.
\begin{equation}
[a(x), a^\dag(x')] = \delta^4(x - x').
\end{equation}
In the model of refs. \cite{BedinghamRSRM,BedinghamRSRD}, the nonstandard field couples to the standard degrees of freedom in such a way that, if the  state on some hyperplane is the ground state of all fields, including the nonstandard field, it remains  that state on all later hypersurfaces, but excitations of the standard fields give rise to excitations of the nonstandard field, which is then subject to collapse dynamics.  Introduction of nonstandard degrees of freedom   may appear to some to be an unmotivated, \emph{ad hoc} move. We have shown that a move of this sort is necessary, for a relativistic dynamical collapse theory formulated within the framework of quantum dynamical semi-groups.

\end{document}